\documentclass[conference]{IEEEtran}
\IEEEoverridecommandlockouts
\usepackage[bookmarks=false]{hyperref}
\usepackage{cite}
\usepackage{amsmath,amssymb,amsfonts}
\usepackage{algorithmic}
\usepackage{graphicx}
\usepackage{textcomp}
\usepackage{xcolor}
\usepackage{enumitem}

\usepackage{multirow}
\usepackage{array}
\usepackage{makecell}
\usepackage{booktabs}
\usepackage{placeins}

\usepackage{stfloats}

\usepackage{eso-pic}

\def\BibTeX{{\rm B\kern-.05em{\sc i\kern-.025em b}\kern-.08em
    T\kern-.1667em\lower.7ex\hbox{E}\kern-.125emX}}
    
\begin{document}

\AddToShipoutPictureFG*{%
  \AtPageUpperLeft{%
    \raisebox{-0.8cm}{%
      \hspace{6cm}%
      \footnotesize\textbf{Presented at IEEE International Conference on Communications ICC 2026}
    }%
  }%
}

\makeatletter
\newcommand{\linebreakand}{%
  \end{@IEEEauthorhalign}
  \hfill\mbox{}\par
  \mbox{}\hfill\begin{@IEEEauthorhalign}
}
\makeatother

\title{Burst Aware Forecasting of User Traffic Demand in LEO Satellite Networks\\
}

\author{\IEEEauthorblockN{Yekta Demirci\IEEEauthorrefmark{1},
Guillaume Mantelet\IEEEauthorrefmark{2},
Stéphane Martel\IEEEauthorrefmark{2},
Jean-François Frigon\IEEEauthorrefmark{1},
Gunes Karabulut Kurt\IEEEauthorrefmark{1}}
\IEEEauthorblockA{\IEEEauthorrefmark{1}Poly-Grames Research Center, Department of Electrical Engineering, Polytechnique Montréal, QC, Canada}
\IEEEauthorblockA{\IEEEauthorrefmark{2} Satellite Systems, MDA Space, Canada}
}

\maketitle

\begin{abstract}

In Low Earth Orbit (LEO) satellite networks, Beam Hopping (BH) technology enables the efficient utilization of limited radio resources by adapting to varying user demands and link conditions. Effective BH planning requires prior knowledge of upcoming traffic at the time of scheduling, making forecasting an important sub-task. Forecasting becomes particularly critical under heavy load conditions where an unexpected demand burst combined with link degradation may cause buffer overflows and packet loss. To address this challenge, we propose a burst aware forecasting solution. This challenge may arise in a wide range of wireless networks; therefore, the proposed solution is broadly applicable to settings characterized by bursty traffic patterns where accurate demand forecasting is essential. Our approach introduces three key enhancements to a transformer architecture: (i) a distance from the last burst embedding to capture burst proximity, (ii) two additional linear layers in the decoder to forecast both upcoming bursts and their relative impact, and (iii) use of an asymmetric cost function during model training to better capture burst dynamics. Empirical evaluations in an Earth-fixed cell under high-traffic demand scenario demonstrate that the proposed model reduces prediction error by up to 94$\%$ at a one-step horizon and maintains the ability to accurately capture bursts even near the end of longer prediction horizons following Mean Square Error (MSE) metric. 
\end{abstract}


\section{Introduction}

Low Earth Orbit (LEO) satellite constellations have emerged as a promising solution to complement existing terrestrial networks. The deployment of terrestrial networks is often infeasible in sparsely populated and hard to reach regions. On top of this, the world is experiencing significant inequality in broadband access. This digital divide is expected to increase social inequality as individuals in underserved regions are unable to benefit from remote education and work services \cite{yaacoub2020key}. LEO constellations are strong candidates to help close this divide and complement terrestrial networks via their ability to provide global coverage. Moreover, LEO High Throughput Satellites (HTS) have achieved significant capacity improvements in recent years \cite{yahia2024evolution}, further strengthening their role in delivering broadband service to underserved and non-covered areas. All in all, LEO constellations are expected to become a key component of the 6G ecosystem, complementing terrestrial networks to enable seamless worldwide coverage \cite{azari2022evolution}.

\begin{figure}[t]
\centerline{\includegraphics[width=1\linewidth]{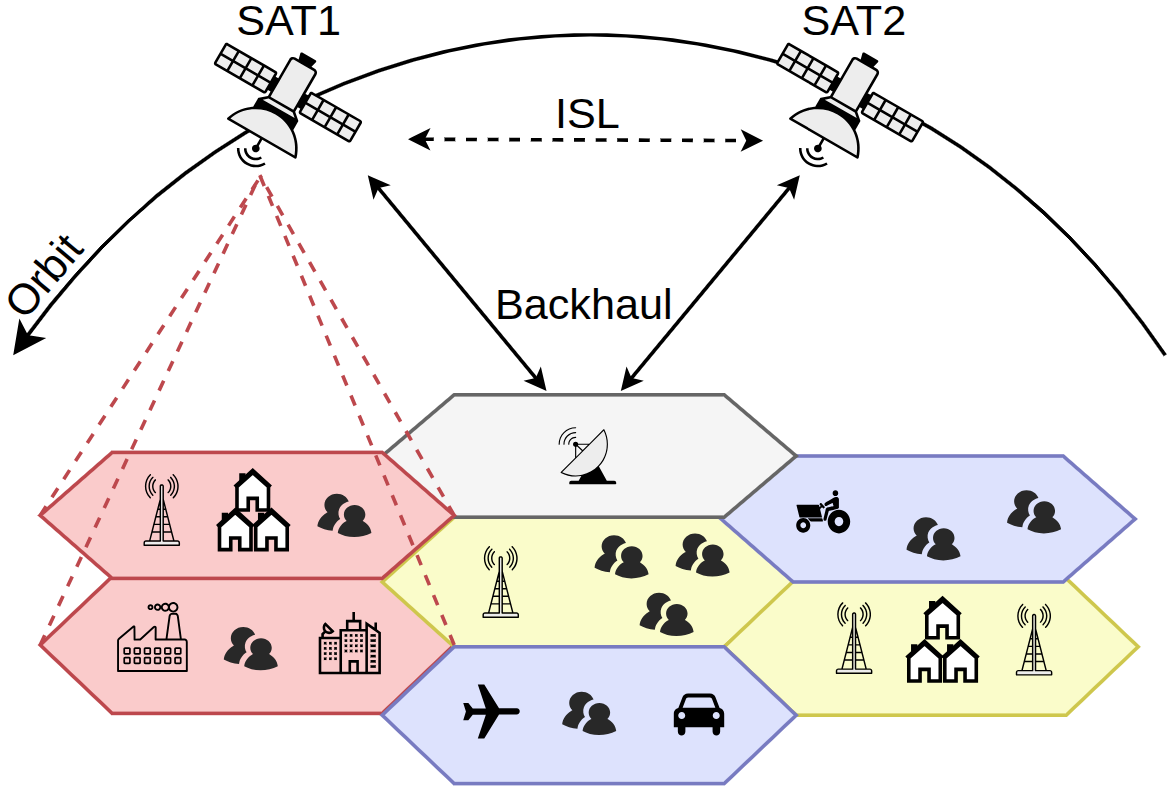}}
\caption{SAT1 serves two user cells experiencing high downlink demand. As SAT1 maintains a stable backhaul connection, packets continue to arrive steadily. Downlink user cells may experience sudden traffic bursts combined with channel quality degradation, which can lead to buffer congestion and even packet loss. }
\label{fig:setting}
\end{figure}

Given the potential of LEO satellite constellations, efficient spectrum and power management remain central challenges where beam hopping emerges as a key enabler. Unlike conventional broadband multi-beam systems which rely on continuously illuminated beams regardless of the demand and the channel conditions, beam hopping introduces temporal and spatial flexibility by selectively activating subsets of beams during specific dwell times. By dynamically adjusting time–space illumination patterns, more power and spectrum can be allocated to higher demand areas. However, effective beam hopping scheduling depends on accurate forecasting of user demand and channel conditions\cite{demirci}. These forecasts become particularly critical under heavy load conditions, where a sharp increase in demand combined with link degradation may cause buffer overflows and packet losses. To address this challenge, we propose novel enhancements to transformer-based architectures to enable burst aware forecasting.

In this work, we consider a LEO satellite constellation where each satellite is equipped with a regenerative payload, providing broadband services to several Earth-fixed cells. The forward-link (downlink) traffic demand in each cell may vary due to different number of active users and heterogeneous service requirements, resulting in non-uniform aggregated demand across cells. A snapshot of this setting is shown in Fig.~\ref{fig:setting}, where Satellite 1 (SAT1) may face high traffic loads in its served cells. SAT1 can connect to the backhaul either directly or through another satellite along its orbital trajectory via an inter-satellite link (ISL). In these cases, SAT1’s buffers may become congested when propagation channel degradation coincides with a sudden burst in downlink demand, as backhaul traffic continues to flow uninterruptedly. This highlights the necessity of forecasting mechanisms to anticipate and manage the upcoming bursts. Similar scenarios can arise across many wireless networks; therefore, the proposed solution is not limited to LEO satellite constellations but can also be applied to a variety of settings where traffic demand is bursty and accurate forecasting is essential.

Considering the existing literature, several works have investigated sudden surges in time series across different domains, ranging from networking to finance. As these surges may represent anomalies, the focus often revolves around early detection \cite{wang2023real, cheng2024burstdetector}. Alternatively, some studies aim at forecasting the underlying time series values during surge events, which is also the focus of our work. Depending on the time granularity of the series and the persistence of the surge, different terms such as peak, spike, or burst are used. For instance, in \cite{zhang2023unlocking}, the authors propose a transformer-based solution to forecast upcoming peaks. However, the work lacks a formal definition of what constitutes a peak, assuming them to be pre-identified, and focuses solely on peak forecasting. In \cite{tat2025}, the authors proposed a transformer-based solution to forecast peaks in a retailer dataset in order to anticipate upcoming demand surges. However, the underlying data exhibits distinct characteristics, including static features, which limit its direct comparability to our setting. In \cite{ding2019modeling}, the authors referred to sudden surges as extreme events after introducing an increase indicator, and then proposed a Gated Recurrent Unit (GRU) to forecast them. However, the use of GRUs is impractical for multi-step forecasting due to their sequential recurrence. Thus, there remains a gap for forecasting solutions that predict the underlying time series while carefully accounting for expected bursts.

Given this, our main contributions are as follows:

\begin{enumerate}[label=C\arabic*]
\item We propose a burst-aware transformer that introduces three novel enhancements applicable to any time series forecasting transformer with an encoder and a generative style decoder architecture: (i) a distance from the last burst embedding to capture burst proximity, (ii) two additional linear layers in the decoder to predict whether a burst is expected and adjust the forecast accordingly, and (iii) an asymmetric loss function applied only to sequences containing burst events during model training.

\item We demonstrate that the proposed solution not only forecasts bursts more accurately than the baseline models but also retains this ability across multiple prediction lengths, even when bursts occur near the end of the horizon.

\item We provide an ablation study to quantify the individual impact of each proposed enhancement on burst and overall forecasting performance.

\item We present a comprehensive empirical evaluation demonstrating that our model consistently outperforms a baseline transformer across different prediction lengths, achieving up to a 94$\%$ reduction in burst prediction error for one-step-ahead forecasting following Mean Square Error (MSE) metric.
\end{enumerate}
 
\section{Background} \label{background}

\subsection{On the Notion of Bursts}
Although the notion of a burst may vary across different works, 
we adopt a formal definition inspired by \cite{palshikar2009simple}. 
Let $X = (x_0, x_1, \dots, x_{N-1}) \in \mathbb{R}^N$ denote a vector of $N$ values. 
We define the set $B$ to contain the indices at which bursts occur:
\begin{equation}
    B = \{ i \in P :a_i-\mu_P> h \sigma_P, x_i-\mu_X>h\sigma_X  \} 
\label{eq:burst}
\end{equation}

where $h$ is a given constant, $\mu_X$, $\sigma_X$ are the mean and Standard Deviation (SD) of $X$ and $\mu_P$, $\sigma_P$ are the mean and SD of the set $P$:
\begin{equation}
    P=\{j : a_j>0\}, \quad j\in\{k...N-k\},
\end{equation} where $k$ is a given constant and $a_j$ is defined as the following for a given constant $k$:
\begin{equation}
    a_j = x_j -\frac{1}{2k}\left( \sum_{r=1}^k x_{j-r} + \sum_{r=1}^k x_{j+r} \right)\
\label{eq:score}
\end{equation}

\subsection{Second Order Self Similar Traffic Model} \label{traffic}

To characterize user demand within fixed Earth cells, a traffic model or public data is required. However, to the best of our knowledge, no publicly available LEO satellite dataset exists. Motivated by this, we study demand characteristics in satellite networks through established traffic models. Demand is aggregated across users and receivers within a cell, where each user may generate multiple streams. Broadband services are predominantly IP-based, and demand fluctuates sharply over short intervals, producing bursty patterns, suggesting a self-similar traffic model. Three empirical studies confirm this property: Leland et al. analyzed Bellcore LAN Ethernet packets and showed consistent self-similarity across datasets\cite{leland2002self}. Paxson et al. examined WAN traces with up to 3.7M TCP connections, showing that Poisson models fail while traffic exhibits asymptotic self-similarity \cite{paxson}. Crovella et al. analyzed WWW requests from 39 student devices at 10 ms resolution, again confirming self-similar patterns \cite{crovella}. Therefore, we adopt a second-order self-similar traffic model in this work.

The degree of self-similarity can be measured using the Hurst parameter, which describes the burstiness of a given dataset. Following their formulation, let $X = (X_t : t = 0,1,2,\dots)$ denote a covariance-stationary stochastic process with mean $\mu$, variance $\sigma^2$, and autocorrelation function $r(k)$ for $k \geq 0$. For $0 < \beta < 1$, the autocorrelation function is assumed to take the form:
\begin{equation}
r(k) \sim k^{-\beta} L(t) \text{ , } k \rightarrow \infty.
\label{eq:first}
\end{equation}
$L(t)$ is a slowly varying function at infinity for all $x>0$
\begin{equation}
    \text{lim}_{t\rightarrow \infty} \frac{L(tx)}{L(t)} = 1.
\end{equation}
Let $X^{(m)} = (X_k^{(m)} : k = 1,2,3,\dots)$ denote a new covariance-stationary time series obtained by averaging $X$ over non-overlapping blocks of size $m$. More formally:
\begin{equation}
    X_k^{(m)} = \frac{1}{m}(X_{km-m+1}+...+X_{km}), \text{ }k\geq1.
\end{equation}

The process $X$ is considered as (exactly) second-order self-similar if both \eqref{eq:third} and \eqref{eq:fourth} hold.  
It is considered as (asymptotically) second-order self-similar if \eqref{eq:fifth} holds for all sufficiently large $k$:
 
 \begin{equation}
\text{var}(X^{m})=\sigma^2m^{-\beta} \text{ , } \forall m
 \label{eq:third}
 \end{equation}
 \begin{equation}
     r^{(m)}(k)=r(k) \text{ , } k\geq0
\label{eq:fourth}
 \end{equation}
\begin{equation}
    r^{(m)}(k) \rightarrow r(k) \text{ , } m \rightarrow \infty. 
\label{eq:fifth}
\end{equation}
If a process is second-order self-similar, its degree of self-similarity is parametrized by the Hurst parameter $H$, where $H = 1 - \beta/2$.

\begin{figure*}[t]
\centerline{\includegraphics[width=1\textwidth]{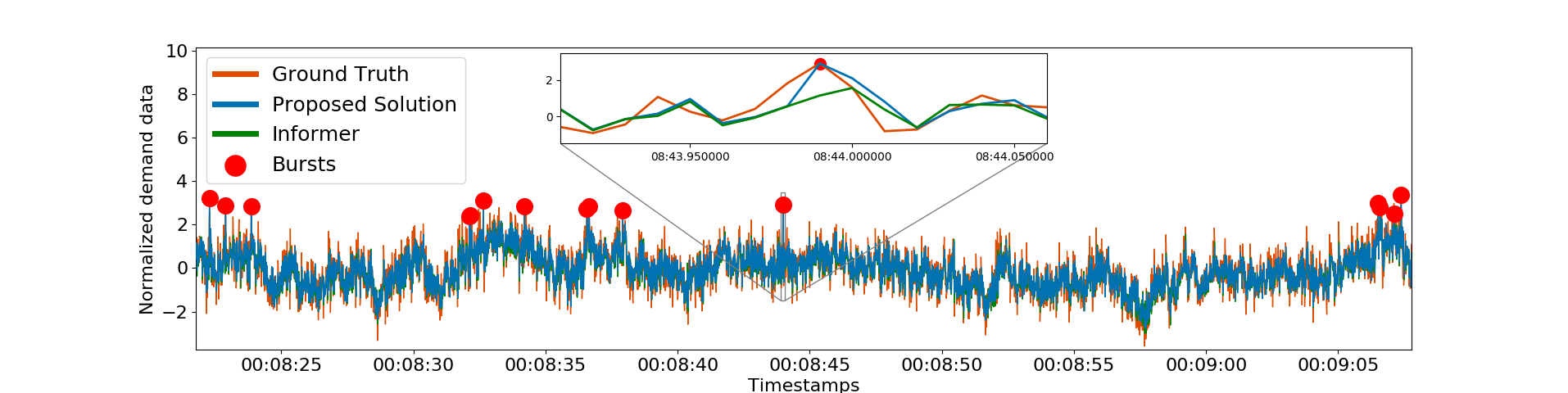}}
\caption{Snapshot of normalized data with one-step-ahead forecasts from the Proposed Solution, Informer, with a zoomed view around an observed burst.}
\label{fig:snapshot}
\end{figure*}

Given this traffic model, user demand data can be synthetically generated using \textit{ON} and \textit{OFF} packet trains \cite{taqqu1997proof}. 
Suppose there are $M$ independent and identically distributed (i.i.d.) sources, each generating a binary time series $\{W(t), t \geq 0\}$, 
where $W(t) = 1$ indicates the presence of a packet at time $t$ and $W(t) = 0$ otherwise. 
The superposition of these $M$ sources, rescaled by a factor $T$, gives the total packet count at time $t$ as:

\begin{equation}
    W^{*}_{M}(Tt) = \int_{0}^{Tt}( \sum_{m=1}^M W^{(m)}(u) )du.
\label{eq:superpose}
\end{equation}

Assume $f_i(x)$ denotes the probability density function of the \textit{ON}/\textit{OFF} periods, 
with $i=1$ corresponding to \textit{ON} and $i=2$ corresponding to \textit{OFF}. 
Let $F_i(x) = \int_0^x f_i(u)\,du$ denote the cumulative distribution function, 
and define its complement as $F_{ic}(x) = 1 - F_i(x)$. 
The mean and variance are given respectively by
\[
\mu_i = \int_0^\infty x f_i(x)\,dx, 
\quad \sigma_i^2 = \int_0^\infty (x - \mu_i)^2 f_i(x)\,dx.
\]
Assume that, as $x \to \infty$, either condition \eqref{eq:cond1} holds 
or $\sigma_i^2 < \infty$ holds for $i \in \{1,2\}$, 
where $l_i > 0$ is a constant and $L_i(x)$ is a slowly varying function at infinity satisfying
\[
\lim_{x \to \infty} \frac{L_i(tx)}{L_i(x)} = 1, \quad \text{for all } t > 0.
\]
Then
\begin{equation}
    F_{ic}(x)\sim l_ix^{-a_i}L_i(x) \text{ , } 1<a_i<2.
\label{eq:cond1}
\end{equation}
Here, $a$ denotes the shape parameter of the respective distribution. 
For large values of $M$ and $T$, the stochastic process defined in \eqref{eq:superpose} 
can be appropriately normalized to $\{\sigma_{\text{lim}} B_H(t), \, t \geq 0\}$ \cite{taqqu1997proof}. 
This normalization follows from the limits in \eqref{eq:limits}, where $\mathcal{L}\!\lim$ denotes convergence in the sense of finite-dimensional distributions. For a detailed proof, interested readers are referred to the cited work:
\begin{equation}
    \mathcal{L} \lim_{T \rightarrow \infty} \mathcal{L} \lim_{M \rightarrow \infty} \frac{\left(W^*_M(Tt)-TM\frac{\mu_1}{\mu_2+\mu_2}t\right)}{T^HL^{1/2}(T)M^{1/2}}=\sigma_{lim}B_H(t).
\label{eq:limits}
\end{equation}
$\sigma_{lim}$ is a finite positive constant and $B_H$ is fractional Brownian Motion, which is a Gaussian Process with stationary increments that is self-similar. $B_H$ has the following covariance function where $H$ is the Hurst parameter:
\begin{equation}
    \text{cov}(B_H(s),B_H(t))=\frac{1}{2}\left\{s^{2H}+t^{2H}-|s-t|^{2H}\right\}.
\label{eq:cov}
\end{equation}

The relationship between the shape and Hurst parameters is given by $H = (3 - a_{\min})/2$. 
An important observation, as pointed out in \cite{taqqu1997proof}, is that the source with the highest $H$ (or equivalently the smallest $a$) ultimately dominates as $T \to \infty$. 
In other words, only the source corresponding to the smallest $a$ parameter among the $M$ sources is relevant. By employing \textit{ON}/\textit{OFF} packet trains, we can generate synthetic data for the considered service time intervals with high granularity 
(\textit{e.g.}, 10 milliseconds), as illustrated in Fig.~\ref{fig:snapshot}.

\begin{figure*}[t]
\centerline{\includegraphics[width=0.7\textwidth]{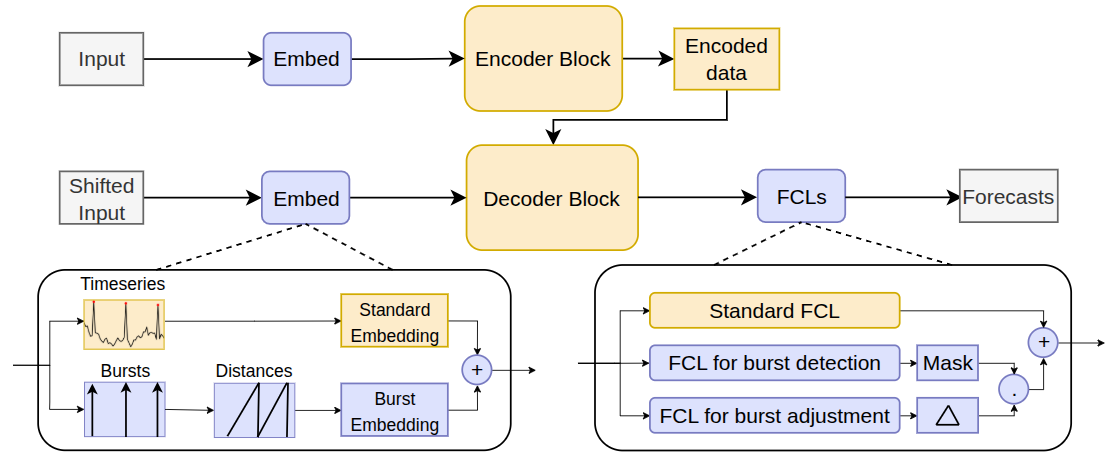}}
\caption{Proposed system architecture, with blue boxes marking enhancements to the forecasting transformer.}
\label{fig:system}
\end{figure*}

\subsection{ARIMA and FARIMA for Statistical Forecasting}\label{arima}
The AutoRegressive Integrated Moving Average (ARIMA) model is a linear approach widely used in statistical time series analysis \cite{beran2013long}. An ARIMA$(p,d,q)$ process, following the notation of Box and Jenkins \cite{beran2013long}, can be defined by \eqref{eq:arima}, 
where $(1-B)^d$ denotes the differencing operator with $d \in \{1,2,3,\dots\}$.
\begin{equation}
    (1-B)^dZ_t=Y_t \text{, } t\geq1.
\label{eq:arima}
\end{equation}
$Y_t$ can be modeled with Autoregressive Moving-Average (ARMA). An ARMA$(p,q)$ process is given by \eqref{eq:arma} where $\varepsilon_t$ is assumed to be i.i.d. with zero mean and finite variance $\sigma_{\varepsilon}^2$. $\phi(k)=1-\sum_{j=1}^p\phi_jk^j$ and $\psi(k)=\sum_{j=0}^q\psi_jk^j$ are polynomials with no common roots and all roots being outside the unit circle \cite{beran2013long}:
\begin{equation}
    \phi(B)Y_t=\psi(B)\varepsilon_t.
\label{eq:arma}
\end{equation}

\subsection{Burst Aware Transformer}
The proposed modifications are applicable to any transformer-based forecasting architecture employing an encoder and a generative decoder.
The first enhancement introduces a burst embedding, which encodes the temporal distance from the last observed burst. This distance grows linearly with time and resets to zero at a burst. To emphasize short intervals, we apply a logarithmic transform \cite{lima2023large}, 
$\log(1+d_l)$, where $d_l$ denotes the distance from the last burst. 
The transformed values are then projected through a linear layer to construct the burst embedding matrix, which is subsequently added to the standard embedding matrices.

As the second enhancement, we introduce two additional Fully Connected Layers (FCLs) at the end of the decoder block, 
in addition to the standard one(s). 
The first added layer predicts whether a burst is expected in the upcoming prediction window, 
using a sigmoid activation function $s(\cdot)$ defined as
\begin{equation}
    s(x) = \frac{1}{1 + e^{-x}}.
\end{equation}

The output of $s(\cdot)$, representing the probability of observing a burst in the prediction window, is used as a mask. A Hadamard product is then applied between this mask and $\Delta$ where $\Delta$ denotes the offset between the regular forecasts and the expected burst. The value of $\Delta$ is obtained from the output of the second FCL, passed through a smooth approximation function $f(\cdot)$ defined as
\begin{equation}
    f(x) = \ln(1 + e^x).
\end{equation}

As the third enhancement, we introduce an asymmetric cost function $L_{asym}(\cdot)$ to be used during training. Let $y_i$ denote the forecast values for a prediction length of $PL$, where $i \in \{1, \dots, PL\}$. 
When computing the loss for this prediction horizon, errors corresponding to burst points are penalized asymmetrically, encouraging the model to better learn peak forecasting.

\begin{equation}
    {L}_{\text{asym}}(y_i,\hat{y}_i,y_i^p)
    = \frac{1}{N} \sum_{i=1}^{N} 
      \left( 1 + \gamma \, y^{\text{p}}_i \right) 
      \left( \hat{y}_i - y_i \right)^{2},
\label{eq:asym}
\end{equation}
where $\gamma$ is a constant, $\hat{y}_i$ denotes the forecast values during training, 
and $y_i^p$ is a binary indicator specifying whether the sample is labeled as a burst. For the first added FCL, we used weighted binary cross-entropy loss and for the second FCL we used mean absolute loss applied only where there is a peak. All loss terms are weighted by 0.33 and summed to form the final training loss used in backpropagation.
\section{Implementation \& Numerical Results}

\begin{table}[b!]
\centering
\caption{Implementation Parameters}
\scriptsize
\begin{tabular}{|l|l|}
\hline
\multicolumn{2}{|l|}{\textbf{Data Generation and Burst Labeling}} \\ \hline
$M$: Number of terminals Eq.~\ref{eq:superpose}& 750 \\ \hline
$a_i$: Pareto shape $\forall i$ Eq.~\ref{eq:cond1}& 0.98 \\ \hline
$k$: Window size Eq.~\ref{eq:burst} & 128 \\ \hline
$h$: SD multiplier Eq.~\ref{eq:burst}& 2.5 \\ \hline
\multicolumn{2}{|l|}{\textbf{Transformer Model and Training}} \\ \hline
$d$: Model length & 512\\ \hline
$n_{head}$: Number of multi heads & 8\\ \hline
Input sequence length & 128 \\ \hline
Input label length & 64 \\ \hline
$P_L$ Prediction length & \{1,12,24,48 \}\\ \hline
\# Encoders \& decoders & 2 \& 1 \\ \hline
Number of epochs & 15 \\ \hline
Batch size & 128 \\ \hline
$\gamma$: Penalizing constant Eq.~\ref{eq:asym} & 5 \\ \hline
\multicolumn{2}{|l|}{\textbf{(F)ARIMA Models}} \\ \hline
$(p,q)$: ARMA order Eq.~\ref{eq:arma} & $(2,0)$ \\ \hline
\end{tabular}
\label{tab:sim_params}
\end{table}

\renewcommand{\arraystretch}{1.5} 
\setlength{\tabcolsep}{4pt}
\newcolumntype{M}[1]{>{\centering\arraybackslash}m{#1}}

\begin{table*}[!t]
\centering
\caption{Forecasting Accuracy by Horizon and Prediction Length}
\small

\begin{tabular}{
!{\vrule width 1.1pt}c|c !{\vrule width 1.1pt} c|c!{\vrule width 1.1pt}c|c !{\vrule width 1.1pt}c|c!{\vrule width 1.1pt}c|c!{\vrule width 1.1pt}}
\Xhline{1.1pt}
\multicolumn{2}{ !{\vrule width 1.1pt}  c!{\vrule width 1.1pt}}{Prediction Length} & \multicolumn{2}{c!{\vrule width 1.1pt}}{1} & \multicolumn{2}{c!{\vrule width 1.1pt}}{12} & \multicolumn{2}{c!{\vrule width 1.1pt}}{24} & \multicolumn{2}{c !{\vrule width 1.1pt}}{48} \\ \hline

\multicolumn{2}{!{\vrule width 1.1pt}c!{\vrule width 1.1pt}}{Overall $\&$ Bursts} & O & B & O & B & O & B & O & B \\ 

\Xcline{1-10}{1.1pt}

\multirow{4}{*}{\rotatebox{90}{Models}}

& Proposed Solution & \textbf{0.3296} & \textbf{0.1095} & 0.5981 & \textbf{0.7462} & \textbf{0.6436} & \textbf{0.8129} & \textbf{0.6999} & \textbf{0.8302} \\ \cline{2-10}

& Informer & 0.3341 & 1.9929 & \textbf{0.5888} & 1.3300 & 0.6482 & 1.1943 & 0.7118 & 1.1568 \\ \cline{2-10}
                     
& FARIMA & 0.3393 & 1.8565 & 0.5910 & 1.2897 & 0.6523 & 1.1051 & 0.7186 & 1.0165 \\ \cline{2-10}
                     
& ARIMA & 0.4031 & 2.4016 & 0.7051 & 1.5726 & 0.7373 & 1.2669 & 0.7766 & 1.1113 \\ 
\Xhline{1.1pt}
\end{tabular}

\vspace{0.1em}

\label{tab:results}
\end{table*}

\subsection{Data generation}
Using \eqref{eq:superpose}, we consider $M$ terminals with active downlink connections to a satellite. Following condition \eqref{eq:cond1}, the Pareto distribution is employed as the long-tail distribution, with the shape parameter set to $a = 1.04$, yielding a Hurst parameter of $H = 0.98$ and thus producing bursty traffic, 
as inspired by \cite{leland2002self}. 
Each terminal generates demand sampled from a normal distribution with mean 1~Mbps and standard deviation 0.05~Mbps. 
The time granularity is 10~ms, meaning that each sample corresponds to the demand within a 10~ms interval. Based on this model, we generate 60,000 samples to create a high-demand traffic dataset. Finally, burst points are labeled using \eqref{eq:burst} with parameters $k = 128$ and $h = 2.5$.

\begin{figure*}[b]
\centerline{\includegraphics[width=1\textwidth]{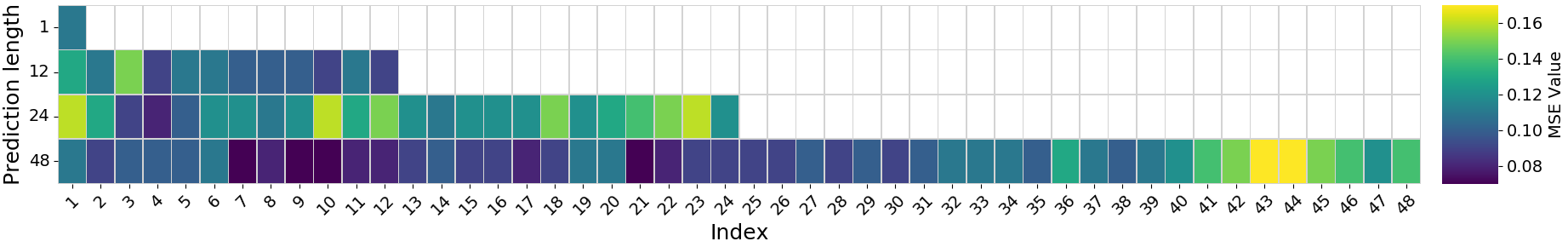}}
\caption{Heatmap showing MSE for burst samples, measured at forecast indices where bursts occur, across different prediction lengths.}
\label{fig:idx}
\end{figure*}


\subsection{Transformer model}
We implemented our proposed modifications on the Informer model \cite{zhou2021informer}, 
a transformer architecture that employs multi-head ``Prob-sparse'' self-attention. 
Nevertheless, the modifications are general and can be applied to any transformer model 
with an encoder and a decoder designed for generative forecasting. The ``Prob-sparse'' attention mechanism $A(Q,K,V)$ reduces the quadratic time complexity of vanilla transformers by selecting only a subset of queries and keys, rather than utilizing all of them \cite{zhou2021informer}.

\begin{equation}
Attention(Q,K,V)=softmax(\frac{\bar{Q}K^T}{\sqrt{d_k}}V).
\end{equation}
Let $Q$, $K$, and $V$ denote the Query, Key, and Value matrices, respectively, 
with $Q \in \mathbb{R}^{L_Q \times d}$, $K \in \mathbb{R}^{L_K \times d}$, 
and $V \in \mathbb{R}^{L_V \times d}$. 
Here, $L_Q = L_K = L_V$ denotes the input sequence length, $d$ is the model dimension, $n_{\text{head}}$ is the number of attention heads, and $d_k = d / n_{\text{head}}$ is the dimension of each head. The ``Prob-sparse" self attention uses a matrix of $\bar{Q}$ containing the Top-$\bar{L}_Q$ queries of $Q$ under the sparsity  measurement of $M(q,K)$. $\bar{L}_Q$ is defined as $(c \ln L_Q)$ where $\bar{L}_Q =\bar{L}_K$.

For each query $q_i \in \mathbb{R}^d$ and $k_j \in \mathbb{R}^d$ in the keys set $K$, $M(q_i,K)$ is bounded as follows:
\begin{equation}
    \ln L_K \leq M(q_i,k) \leq \bar{M}(q_i,K)+ \ln L_K
\end{equation}
where $\bar{M}(q_i,K)$ is defined as:
\begin{equation}
    \bar{M}(q_i,k)=\underset{j}{{\text{max}}} \left\{ \frac{q_i k_j^{\mathsf{T}}}{\sqrt{d}} \right\} - \frac{1}{L_K}\sum_{j=1}^{L_K}\frac{q_ik_j^{\mathsf{T}}}{\sqrt{d}} 
\end{equation}
As a result, the ``ProbSparse'' attention achieves a time complexity of $\mathcal{O}(L_Q \log L_Q)$, in contrast to the quadratic complexity of full attention. Our proposed modifications to the Informer model are available in the project's GitHub repository.\footnote{\url{https://github.com/YektaDemirci/burstAwareForecasting}}.

\subsection{Simulation results}

We considered 4 different prediction length of 1, 12, 24 and 48. We employed the \textit{ARIMA} method from the \textit{statsmodels} library \cite{seabold2010statsmodels} in \textit{Python} and the \textit{arfima} package \cite{veenstra2013persistence} in \textit{R} for the ARIMA and FARIMA models, respectively. For both models, the parameters $p$ and $q$ were set to 2 and 0. The transformer models were trained on an NVIDIA RTX A2000 GPU with 12 GB of memory, and the forecasting tasks were subsequently executed on an NVIDIA RTX 500 Ada GPU with 4 GB of memory. We followed MSE metric for the accuracy comparison $MSE(\bar{y},y)=\frac{1}{n}\sum_{i=1}^n(\hat{y}-y)^2$ where $\bar{y}$ represents the forecast and $y$ ground truth.

The forecasting results are summarized in Table~\ref{tab:results}. To evaluate both general and burst specific performance, we provide accuracies under two labels: Overall (O) and Bursts (B). The overall, (O), accuracy reflects performance across all samples, regardless of whether a burst occurs or not within the forecast horizon. Whereas, the bursts, (B), accuracy specifically measures performance on sequences that contain at least one burst sample. As shown in the results, our model provides substantial improvements in burst forecasting. The vanilla Informer and the statistical models fail to capture bursts as it can be seen from the prediction length of 1, whereas our approach achieves a 94$\%$ reduction in error at this horizon. Moreover, even at a prediction length of 48, sequences containing at least one burst are forecasted with a 28$\%$ improvement. Benchmark models often miss peaks with large errors, so even gains limited to peak forecasting translate into better overall accuracy. Therefore, our model also achieves slightly higher overall forecasting accuracy, except at a prediction length of 12.

Furthermore, the position of bursts within the prediction horizon does not significantly affect forecasting accuracy, 
as shown in Fig.~\ref{fig:idx}. Even for prediction length of 48, bursts occurring near the end of the horizon can still be forecasted with MSE values in the range $0.07$–$0.17$.

Finally, we conduct an ablation study, using one-step horizon to quantify the incremental impact of each proposed modification to the transformer: positional embedding (PE), two additional fully connected layers (FCLs), and the asymmetric cost function (ACF). As shown in Table~\ref{tab:last}, each enhancement improves the model’s ability to forecast bursts.

\begin{table}[!t]
\centering
\caption{Forecasting accuracies from the ablation study}
\small 
\resizebox{0.3\textwidth}{!}{
\begin{tabular}{
!{\vrule width 1.1pt}c|c !{\vrule width 1.1pt} c|c!{\vrule width 1.1pt}}

\Xhline{1.1pt}
\multicolumn{2}{ !{\vrule width 1.1pt}  c!{\vrule width 1.1pt}}{Prediction Length} & \multicolumn{2}{c!{\vrule width 1.1pt}}{1} \\ \hline

\multicolumn{2}{!{\vrule width 1.1pt}c!{\vrule width 1.1pt}}{Overall $\&$ Bursts} & O & B\\ 

\hline

\multirow{4}{*}{\rotatebox{90}{Models}}

& Informer (INF) &0.3341 & 1.9929 \\ \cline{2-4}

& INF + PE & 0.3780 & 1.9534 \\ \cline{2-4}
                     
& INF+PE+FCLs & 0.3314 & 0.6754 \\ \cline{2-4}
                     
& INF+PE+FCLs+ACF & \textbf{0.3296} & \textbf{0.1095} \\ \cline{2-4}
\Xhline{1.1pt}
\end{tabular}
}

\label{tab:last}
\end{table}

\section{Conclusion} \label{conc}
In this work, we introduce three novel enhancements to transformer-based time series forecasting models with encoder and generative decoder blocks. The enhancements are a burst embedding, additional decoder layers for burst prediction and adjustment, and an asymmetric loss function to be used during training specifically for samples containing a burst. These enhancements are particularly important under heavy-load conditions, where a sharp increase in demand combined with link degradation may congest buffers and even result in packet loss. By identifying such burst values in advance, we pave the way for preventive strategies. Our approach significantly improves the accuracy of forecasting burst values regardless of where the bursts occur within the prediction horizon, even at long horizons such as 48 steps. Moreover, these improvements in burst forecasting also translate into better overall accuracy, highlighting the importance of explicitly forecasting bursts in demand. As future work, we plan to extend this framework by exploring lighter transformer variants to support scalable deployment in resource-constrained environments, particularly in LEO satellite networks.
\section*{Acknowledgment}
This work was supported in part by MDA Space; in part by the Consortium de Recherche et d’innovation en Aérospatiale au Québec (CRIAQ); and in part by the Natural Sciences and Engineering Research Council of Canada (NSERC).
\bibliographystyle{IEEEtran} 
\bibliography{main}

@ARTICLE{yahia2024evolution,
  author={Yahia, Olfa Ben and others},
  journal={IEEE Com. Surveys \& Tutorials}, 
  title={Evolution of High-Throughput Satellite Systems: A Vision of Programmable Regenerative Payload}, 
  year={2025},
  volume={27},
  number={3},
  pages={1565-1597},
  doi={10.1109/COMST.2024.3450292}}

@book{beran2013long,
  title={Long-Memory Processes: Probabilistic Properties and Statistical Methods},
  author={Beran, J. and Feng, Y. and Ghosh, S. and Kulik, R.},
  isbn={9783642355127},
  series={SpringerLink : B{\"u}cher},
  year={2013},
  publisher={Springer Berlin Heidelberg}
}

@ARTICLE{leland2002self,
  author={Leland, W.E. and Taqqu, M.S. and Willinger, W. and Wilson, D.V.},
  journal={IEEE/ACM Transactions on Networking}, 
  title={On the self-similar nature of {E}thernet traffic (extended version)}, 
  year={1994},
  volume={2},
  number={1},
  pages={1-15},
  doi={10.1109/90.282603}}

@article{taqqu1997proof,
author = {Taqqu, Murad S. and Willinger, Walter and Sherman, Robert},
title = {Proof of a fundamental result in self-similar traffic modeling},
year = {1997},
issue_date = {Apr. 1997},
publisher = {Association for Computing Machinery},
address = {New York, NY, USA},
volume = {27},
number = {2},
issn = {0146-4833},
doi = {10.1145/263876.263879},
journal = {SIGCOMM Comput. Commun. Rev.},
month = apr,
pages = {5–23},
numpages = {19}
}

@inproceedings{zhou2021informer,
  title={Informer: Beyond efficient transformer for long sequence time-series forecasting},
  author={Zhou, Haoyi and Zhang, Shanghang and Peng, Jieqi and Zhang, Shuai and Li, Jianxin and Xiong, Hui and Zhang, Wancai},
  booktitle={Proceedings of the AAAI conference on artificial intelligence},
  volume={35},
  number={12},
  pages={11106--11115},
  year={2021}
}

@article{seabold2010statsmodels,
  title={Statsmodels: econometric and statistical modeling with python.},
  author={Seabold, Skipper and Perktold, Josef and others},
  journal={SciPy},
  volume={7},
  number={1},
  pages={92--96},
  year={2010}
}

@article{lima2023large,
  title={A large comparison of normalization methods on time series},
  author={Lima, Felipe Tomazelli and Souza, Vinicius MA},
  journal={Big Data Research},
  volume={34},
  year={2023},
  publisher={Elsevier}
}

@article{demirci,
  title={Forecasting Self-Similar User Traffic Demand Using Transformers in {LEO} Satellite Networks},
  author={Demirci, Yekta and Mantelet, Guillaume and Martel, St{\'e}phane and Frigon, Jean-Fran{\c{c}}ois and Karabulut Kurt, Gunes},
  journal={IEEE International Conference on Wireless for Space and Extreme Environments (WiSEE)},
  note = {(In press) available at arXiv:2509.10917},
  year={2025}
}

@inproceedings{palshikar2009simple,
  title={Simple algorithms for peak detection in time-series},
  author={Palshikar, Girish and others},
  booktitle={Proc. 1st Int. Conf. advanced data analysis, business analytics and intelligence},
  volume={122},
  year={2009}
}

@ARTICLE{crovella,
  author={Crovella, M.E. and Bestavros, A.},
  journal={IEEE/ACM Transactions on Networking}, 
  title={Self-similarity in World Wide Web traffic: evidence and possible causes}, 
  year={1997},
  volume={5},
  number={6},
  pages={835-846},
  doi={10.1109/90.650143}}

@ARTICLE{paxson,
  author={Paxson, V. and Floyd, S.},
  journal={IEEE/ACM Transactions on Networking}, 
  title={Wide area traffic: the failure of Poisson modeling}, 
  year={1995},
  volume={3},
  number={3},
  pages={226-244},
  doi={10.1109/90.392383}}

@inproceedings{cheng2024burstdetector,
  title={BurstDetector: Real-Time and Accurate Across-Period Burst Detection in High-Speed Networks},
  author={Cheng, Zhongyi and Gao, Guoju and Huang, He and Sun, Yu-E and Du, Yang and Wang, Haibo},
  booktitle={IEEE INFOCOM 2024-IEEE Conference on Computer Communications},
  pages={2338--2347},
  year={2024},
  organization={IEEE}
}

@article{wang2023real,
  title={Real-time spread burst detection in data streaming},
  author={Wang, Haibo and Melissourgos, Dimitrios and Ma, Chaoyi and Chen, Shigang},
  journal={ACM on Measurement and Analysis of Computing Systems},
  volume={7},
  number={2},
  pages={1--31},
  year={2023},
  publisher={ACM New York, NY, USA}
}

@inproceedings{ding2019modeling,
  title={Modeling extreme events in time series prediction},
  author={Ding, Daizong and Zhang, Mi and Pan, Xudong and Yang, Min and He, Xiangnan},
  booktitle={25th ACM SIGKDD international conference on knowledge discovery \& data mining},
  pages={1114--1122},
  year={2019}
}

@article{tat2025,
  author  = {Zhao, Zhiyuan et. al.},
  title   = {{TAT}: Temporal-Aligned Transformer for Multi-Horizon Peak Demand Forecasting},
  journal={ACM SIGCOMM Computer Communication Review},
  note    = {({A}ccepted) KDD 2025 available at arXiv:2507.10349},
}

@inproceedings{zhang2023unlocking,
  title={Unlocking the potential of deep learning in peak-hour series forecasting},
  author={Zhang, Zhenwei and Wang, Xin and Xie, Jingyuan and Zhang, Heling and Gu, Yuantao},
  booktitle={32nd ACM International Conference on Information and Knowledge Management},
  pages={4415--4419},
  year={2023}
}

@ARTICLE{yaacoub2020key,
  author={Yaacoub, Elias and Alouini, Mohamed-Slim},
  journal={Proceedings of the IEEE}, 
  title={A Key 6{G} Challenge and Opportunity—Connecting the Base of the Pyramid: A Survey on Rural Connectivity}, 
  year={2020},
  volume={108},
  number={4},
  pages={533-582},
  doi={10.1109/JPROC.2020.2976703}}

@article{azari2022evolution,
  title={Evolution of non-terrestrial networks from 5{G} to 6{G}: A survey},
  author={Azari, M Mahdi and others},
  journal={IEEE communications surveys \& tutorials},
  volume={24},
  number={4},
  pages={2633--2672},
  year={2022},
  publisher={IEEE}
}

@book{veenstra2013persistence,
  title={Persistence and Anti-Persistence: Theory and Software},
  author={Veenstra, Justin Quinn},
  year={2013},
  publisher={Ph.D dissertation, The University of Western Ontario (Canada)}
}

\end{document}